\documentclass[10pt,letterpaper]{article}
\usepackage{opex3}
\usepackage{amsmath}
\usepackage{nicefrac}

\begin{document}
\title{Multi-type particle layer enhanced absorption for photovoltaics}

\author{Christin David}
	\address{DTU Fotonik, Department of Photonics Engineering, Technical University of Denmark, DK-2800 Kgs. Lyngby, Denmark}
	\email{chrida@fotonik.dtu.dk}
\begin{abstract} 
We discuss regular particle arrays as nanostructured front layers for 3rd generation photovoltaic devices. A rigorous plane-wave method is used to investigate multi-type particle layers combining different radii and configurations. 
The absorbance is enhanced compared to the bare wafer and on mixing particles we demonstrate a broadband boost in the absorbance within the photo-active region. Efficiency enhancement in terms of short circuit current is studied for varying geometries. Si multi-type layers achieved up to 33\% improvement without yet standard anti-reflection coatings. Metal multi-type layers show strong parasitic absorption and boost the absorbance only in narrow wavelength regions and no efficiency enhancement is observed.
\end{abstract}
\bibliographystyle{osajnl}

\section{Introduction}
The trend in solar cell fabrication towards ultrathin wafers, a cost effective and environment-friendly use of materials involves the exploration of additional nanostructured layers to enhance intrinsic quantum efficiencies. Nanotechnology has opened up different routes to reduce front surface reflection, \cite{Branz2009,Liu2012a,Cortes-Juan2013} boost internal absorption through up- and down-conversion \cite{Fischer2010, Yuan2011, Huang2013} and increase light trapping employing particle layers as back reflectors. \cite{Ferry2008, Basch2012, Basch2012a, Brongersma2014}

Both regular and irregular (self-assembled) arrays of both dielectric and plasmonic  particles have been investigated for improving light management in 3rd generation solar cells. \cite{Akimov2010,Spinelli2012,Catchpole2008a,Chen2013,Cui2013} Light harvesting has mostly exploited collective and scattering properties, but also field enhancement effects on the nanoparticle surfaces. While regular particle arrays typically show narrow spectral resonances, randomly distributed particles yield a broader optical response at weaker coupling strengths. \cite{David2013a}

Broadband total absorption can thus also be achieved  in a regular particle array combining different particle sizes and shapes to create a spectral overlap of resonances. This has been demonstrated experimentally for metamaterials in the mid-infrared region \cite{Liu2011, Zhang2011a} and total absorption was also predicted for textured graphene surfaces \cite{Thongrattanasiri2012}. We emphasize that in photovoltaic devices it is not enough to reduce the overall device reflectivity or boost its overall absorption: the absorption of the photo-active region must be enhanced and parasitic absorption (resonant absorption in the textured layer, ohmic heating in metal components etc.) reduced. \cite{Duehring2012,Cortes-Juan2013}
With this huge optimization task at hand, computational nanophotonics is a key ingredient to study various configurations and material compositions to assure a high density of optical modes in the active region of a solar cell. \cite{Han2010,Deceglie2012}

In this article, we investigate regular particle arrays as textured front layers made of unit cells that include more than a single particle. We demonstrate that already using two differently sized nanodisks of the same material can increase the internal absorption, i.e. the absorption at the active device level, corrected by losses in the particle array itself. We study the effect of mixing up to four different particle sizes, including aspects of disk height.
The structure is introduced in the next section and general aspects on the computational procedure are made in section 3.

\begin{figure*}%
\centering
\includegraphics[width=0.5\columnwidth]{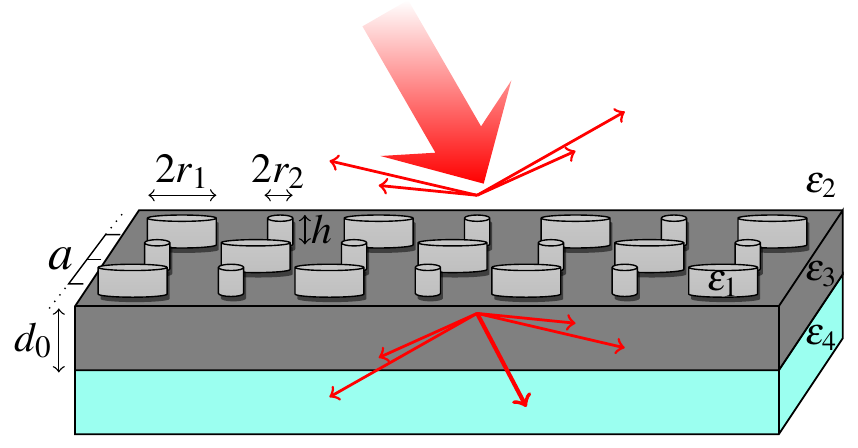} 
\caption{Illustration of the base geometry. A photo-active Si-wafer of thickness $d_0$ and permittivity $\epsilon_3(\omega)$ is placed on a glass substrate with permittivity $\epsilon_4(\omega)$. A photonic crystal structure composed of disk particles with height $h$, periodicity $a$ and radii $r_1, r_2 \dots$ (here two types of Ag disks depicted) and a space-modulated permittivity $\epsilon_1(\vec r,\omega)$ lies on top of the Si-wafer. The outer environment is air ($\epsilon_2=1$) and tabulated data is used for wavelength dependent permittivities.} 
\label{fig1}%
\end{figure*}

\section{Device structure and multi-type layers}

\begin{figure*}%
\centering
\includegraphics[width=\textwidth]{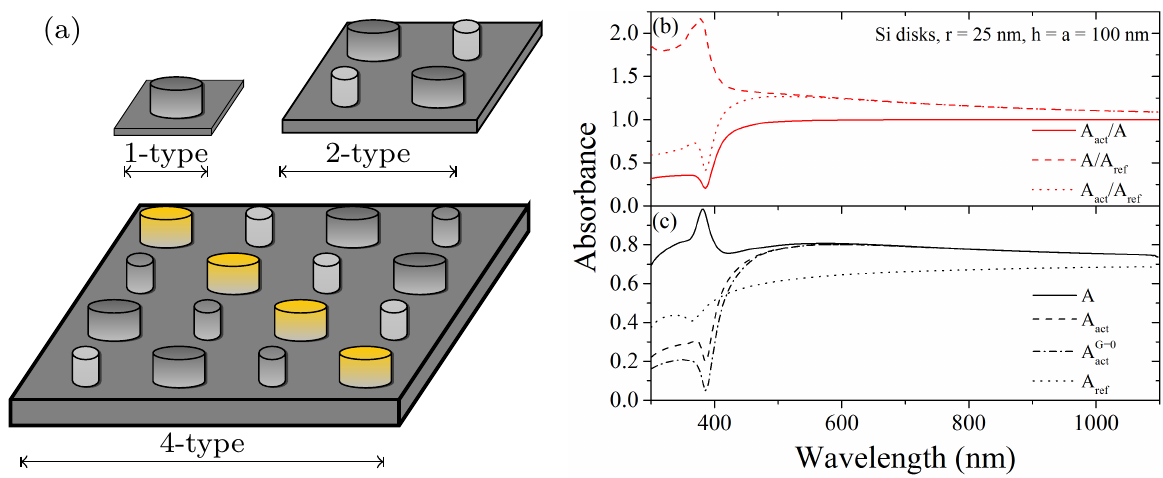}
\caption{(a) Unit cells used in the calculations for multi-type particle layers from the standard 1-type unit cell to an extension to 2 and 4-types unit cells which allows to study particle layers of varying composition and configuration. (b), (c) Calculated quantities are the absorbance of the bare Si reference cell $A_{\rm{ref}}$, the overall device absorption $A$ and the absorption in the photo-active Si-wafer region $A_{\rm{act}}$, i.e. the portion of light absorbed in the region of the reference cell. Typically, we present the optical properties of textured cells relative to the absorbance of the reference cell as in (c). The example spectrum shows these quantities for $r=25$nm Si particle layer with $a=100$nm, $h=100$nm, comparing with the fraction of absorbance arising from only the lowest order of diffraction $A^{G=0}_{\rm{act}}$.}
\label{fig2}%
\end{figure*}

Figure \ref{fig1} illustrates the setup studied: A particle layer with space-modulated permittivity $\epsilon_1(\vec r,\omega)$ is placed on top of a Si wafer (permittivity $\epsilon_3(\omega)$ and constant thickness $d_0=2000\mu$m in order to suppress Fabry-P\'erot like oscillations in the calculated absorption) which is supported on a SiO$_2$ substrate with permittivity $\epsilon_4(\omega)$. Both lower layers are considered to be homogeneous and described by a wavelength-dependent permittivity. The particle layer on top of them scatters the incoming light (air $\epsilon_2=1$, no standard anti-reflection coating used) into a number of diffraction orders for both reflected and transmitted light and thus potentially increases absorptivity in the photo-active, homogeneous Si region underneath by means of an increased optical path length, and to a minor degree through field enhancement effects directly at the interface between the particle layer and the solar cell wafer.
In particular, we want to study regular particle layers that consist not only of the same particles, but different types of disk-shaped particles with equal height $h$ either varying in radius or material and combinations of those. Fig. \ref{fig2}(a) depicts such unit cells, where we use the term 1-type for a unit cell with a single particle, a unit cell with $2\times2$ particles is denoted 2-type and a unit cell with $4\times4$ particles 4-type, since we restrict ourselves to the symmetric cases shown in Fig. \ref{fig2}(a). The modulation of the particle layer is determined by the periodicity $a$ of the unit cell which is also the particle to particle separation for 1-type particle layers.

We exploit here the plane-wave expansion method to calculate the properties of the particle layer and a scattering matrix approach to obtain the overall optical properties of the structure. \cite{Whittaker1999,Liu2012} In order to describe the material properties of the particle layer, a Fourier transform of the real-space, piecewise defined permittivity within the unit cell of area $A$ of the regular structure is made
\begin{align}
	\epsilon(\omega,\vec r)&=\sum_{\vec G}\epsilon_{\vec G} e^{i\vec G \vec R},
\intertext{with reciprocal vectors $\vec G$ and coefficients}
	\epsilon_{\vec G\vec G'}&= \frac{1}{a^2}\int_{A}\epsilon(\omega,\vec r)e^{-i(\vec G-\vec G')\vec R}dA,
\end{align}
which is analytical for 2D circular shapes, see for instance \cite{Guo2013}.
This allows for a straightforward extension to include several particles within a single unit cell.
While the computational effort is increased by a finite summation over the number of particles in the unit cell, one has to keep in mind that the size of the unit cell increases and thus, in order to obtain results of equal quality, a similarly higher number of plane-waves has to be considered at the same time, which is the main drawback in this approach. Note that this approach does allow for asymmetric, including random,\cite{David2013a} particle distributions and an arbitrary assignment of material properties, which is, however, not in the scope of this work.

We study mainly the dielectric Si and the metal Ag taking the permittivity of these materials from tabulated, experimental data. Si has a low refractive index contrast to air and convergence is reached faster, i.e. for a lower number of plane waves taken into account than for metal particle arrays which have large negative permittivity in the visible spectrum. Note furthermore, that averaging over orthogonal polarization directions is not necessary as long as we maintain a symmetric unit cell for the 2D photonic crystal layer as depicted in Fig. \ref{fig2}(a).

\section{Procedure}

We calculate the power flux reflected from the outer (inner) surface in front (behind of) the particle layer and the power flux transmitted into the substrate, normalize to the incoming power flux and obtain from this the overall device absorbance $A$ and the absorbance $A_{\rm act}$ of the photo-active wafer region, which does exclude parasitic (ohmic) absorption in the particle layer itself. This is compared relative to the bare reference cell $A_{\rm ref}$, see Fig. \ref{fig2}(b), for the different configurations discussed in the next section. We give in Fig. \ref{fig2}(c) the absolute values and include for the absorbance $A_{\rm act}$ the contribution of the lowest diffraction order (i.e., where the reciprocal lattice vector $\vec G=0$). Whilst this is the dominant contribution at long wavelengths, a clear error would be made, neglecting higher order contributions.

Two spectral regions can be identified from Fig. \ref{fig2}(c) which are significant in the assessment of such setups. In one case, the absorption in the photo-active region is larger than for the reference cell and close to the overall device absorption. In the other case, while the total device absorption reaches up to 100\%, the absorption within the wafer region decreases below the one of the reference cell. In this latter case, the considered particle layer absorbs most of the light (ohmic heating) which in return is no longer available for the active region of the solar cell. In the graphs shown throughout this work, we concentrate on the absorption in the photo-active region relative to the absorbance of the reference cell $A_{\rm act}/A_{\rm ref}$ as the quantity of main interest.

\begin{figure*}%
\centering
\includegraphics[width=\textwidth]{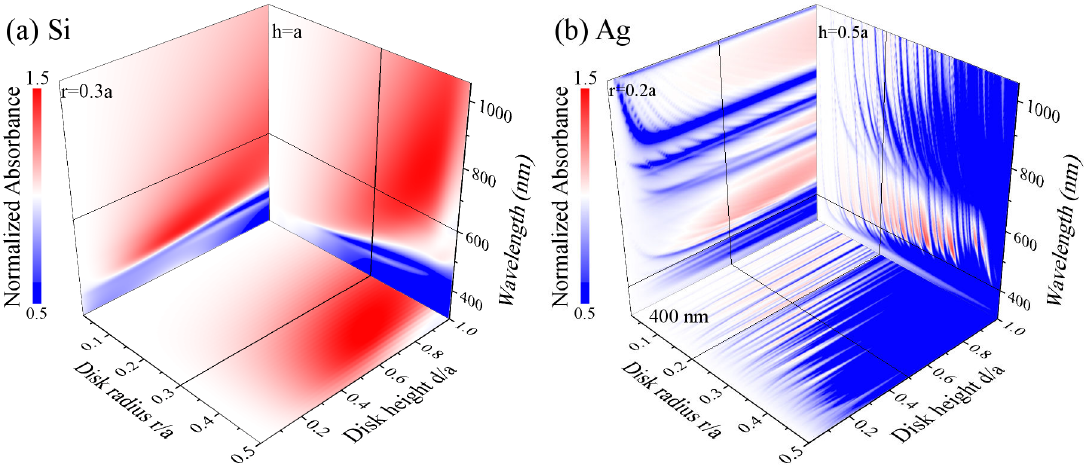}
\caption{Comparison of the absorbance inside the active region of standard Si wafers coated with regular particle arrays using a single particle type (1-type) unit cell, namely (a) Si and (b) Ag, relative to the bare reference cell $A_{\rm{act}}/A_{\rm{ref}}$. The three parts of each graph show the spectra in dependence (a) of disk radius at fixed height $h=100$nm (right), disk height at fixed radius $r=30$nm (left) and for $\lambda=600$nm the mutual dependence on both height and radius (bottom). Similarly, for (b) with $h=50$nm, $r=20$nm and $\lambda=400$nm. The incident light is TM-polarized at normal incidence and the array period is $a=100$nm.}
\label{fig3}%
\end{figure*}

Furthermore, we compute the short circuit current $I_{\rm{sc}}$ for the solar cells with particle arrays relative to the bare reference cell
\begin{align} 
\Delta\eta_{\rm{Isc}}=1-\frac{I_{\rm{sc}}}{I^{\rm{ref}}_{\rm{sc}}} = 1-\frac{\nicefrac{q}{(hc)}\int_{\lambda_0}^{\lambda_N} A_{\rm{act}}(\lambda)\lambda \rm{AM}_{1.5G}(\lambda) d\lambda}{\nicefrac{q}{(hc)}\int_{\lambda_0}^{\lambda_N} A_{\rm{ref}}(\lambda)\lambda \rm{AM}_{1.5G}(\lambda) d\lambda}, \label{eq.Isc}
\end{align}
as well as the average absorbance within the photo-active region in the considered spectral range ($>300$nm)
\begin{align}
\Delta\eta_{\rm{Aav}}=1-\frac{A^{\rm{av}}_{\rm{act}}}{A^{\rm{av}}_{\rm{ref}}} = 1-\frac{\int_{\lambda_0}^{\lambda_N} A_{\rm{act}}(\lambda) d\lambda}{\int_{\lambda_0}^{\lambda_N} A_{\rm{ref}}(\lambda) d\lambda}.
\end{align}
Note that eq. \eqref{eq.Isc} takes the absorption in the photo-active region rather than the total device absorption into account. Assuming an ideal internal quantum efficiency $IQE = EQE/A = 1$ would yield for the external quantum efficiency $EQE=A$ with $A$ being the total absorption including all effects of losses, reflected or back-scattered light in the total device. This is the accessible quantity in an experiment, but here, we can also address the losses within the particle array, and thus have a further correction to the $IQE$, which is the efficiency of carriers collected in the active Si wafer region of the device. Thus, we use directly $A_{\rm act}$ which gives the light absorbed in active region, does account for the losses introduced on employing the particle array and is thus a more suitable choice for a direct quantity to compare with $EQE$.

This gives us two quantitative values to assess the efficiency enhancement found for the different setups on the overall (solar) spectrum. Since these quantities are independent of a specific wavelength, they are ideal to study the influence of various geometrical parameters on them.

\section{Results and discussion}

We give an overview of optical properties of 1-type particle layers in Fig. \ref{fig3} for disk arrays with lattice period $a=100$nm in terms of the absorption in the photo-active region relative to the bare Si wafer absorbance. We study the dielectric material Si and the metal Ag, with permittivities taken from tabulated, experimental data. 
An increase of over 50\% in the absorbance (at specific wavelengths) is seen here, which improves even further for higher disks as known from black Silicon \cite{Branz2009, Liu2012a, Kroll2012}. Note that GaAs produces very similar results to Si comparing spectra of $A_{\rm act}/A_{\rm ref}$ as a function of both disk height and disk radius (not shown).

\begin{figure}%
\centering
\includegraphics{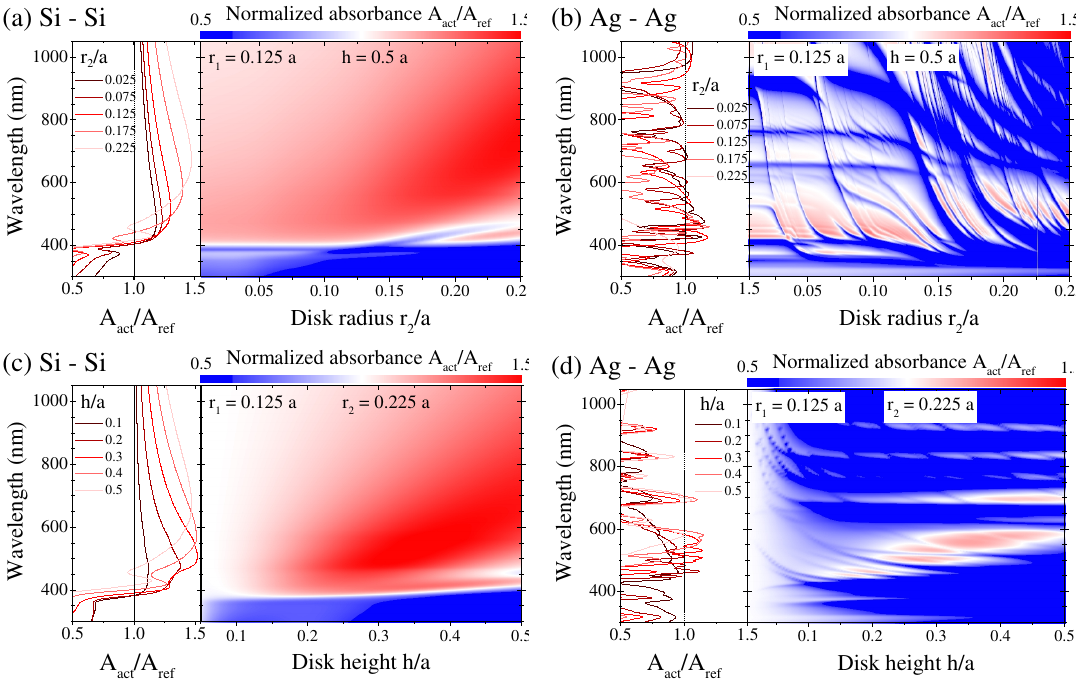}
\caption{Selected disk sizes from 1-type calculations for Si and Ag are combined into 2-type cells. (a), (c) 2-type cells made of Si and (b), (d) Ag disks using (a), (b) different disk radii $r_2$ while keeping constant height $h=100$nm and $r_1=25$nm, and (c), (d) scaling the height for these cells with constant $r_1=25$nm and $r_2=45$nm. The curves plotted on the left hand side of each contour show the spectra that are further evalutated in table \ref{tab.Si2type} for Si.}
\label{fig4}
\end{figure}

The spectral dependence becomes more involved with metal disks, as they show a multitude of sharp resonance features (reflection bands resulting from collective free electron modes). Here, a careful tailoring of geometrical properties is needed to achieve the desired improvement in the absorbance inside the photo-active region. The overall device absorption is readily enhanced mostly via parasitic absorption in the metal particle array. The resulting reduction in overall reflection has prompted plasmonic particle layers ideal for implementation in photovoltaic devices, but the internal quantum efficiency suffers from these losses which cannot be assessed easily in an experiment and are not counterbalanced by adding a standard ARC coating. \cite{Cortes-Juan2013} This can already be seen for the absorbance at low wavelengths for Si in Fig. \ref{fig2}(b) and (c), and we come back to this later on.
The main advantage that can be expected from employing metal disks is to enhance the absorbance at lower wavelengths (below $400$nm in case of Al, between $400\dots600$nm in case of Ag) in a setup using the particle layer as a back reflector for long wavelengths.

\begin{table}%
\centering
\begin{tabular}{c|c|c|c|c||c|c|c}
	Si $r_2$ & Si $r_1$ & $h$     & $\eta_{\rm{Isc}}$ & $\eta_{\rm{Aav}}$ & Si $r_1$ & $\eta_{\rm{Isc}}$ & $\eta_{\rm{Aav}}$ \\\hline
	 5 nm & 25 nm & 100nm & +13\% &  +9\% & 10nm &  +5\% &  +4\% \\
	15 nm & 25 nm & 100nm & +16\% & +11\% & 10nm &  +8\% &  +7\% \\
	25 nm & 25 nm & 100nm & +23\% & +15\% & 10nm & +14\% & +10\% \\
	35 nm & 25 nm & 100nm & +29\% & +19\% & 10nm & +22\% & +17\% \\
	45 nm & 25 nm &  20nm &  +5\% &  +3\% & 10nm &  +4\% &  +3\% \\
	45 nm & 25 nm &  40nm & +17\% & +14\% & 10nm & +15\% & +13\% \\
	45 nm & 25 nm &  60nm & +28\% & +22\% & 10nm & +26\% & +22\% \\
	45 nm & 25 nm &  80nm & +33\% & +24\% & 10nm & +32\% & +23\% \\
	45 nm & 25 nm & 100nm & +32\% & +22\% & 10nm & +31\% & +22\%
\end{tabular}
\caption{Efficiency enhancement calculated from the short circuit current and average absorption relative to the reference cell for Si-Si 2-type cells from Fig. \ref{fig4}(a) and (c) for different particle sizes and heights.} 
\label{tab.Si2type}
\end{table}

From these results, we chose specific disk sizes of Si and Ag that we combine in 2-type particle layers. First, we mix only different radii of the same material in Fig. \ref{fig4} and table \ref{tab.Si2type}, and then we mix Si and Ag disks of equal size in Fig. \ref{fig5}. The main bottleneck in this type of computation is, as mentioned before, that the Fourier transform for larger unit cells needs a higher number $-N\dots N$ of plane waves taken into account to reproduce results for 1-type particle arrays. This directly yields a moderate increase in the computational time $(2N)^2\rightarrow (2mN)^2$, with $m$ the multi-type number (square root of particle number in the unit cell).

\begin{figure}%
\centering
\includegraphics[width=0.45\textwidth]{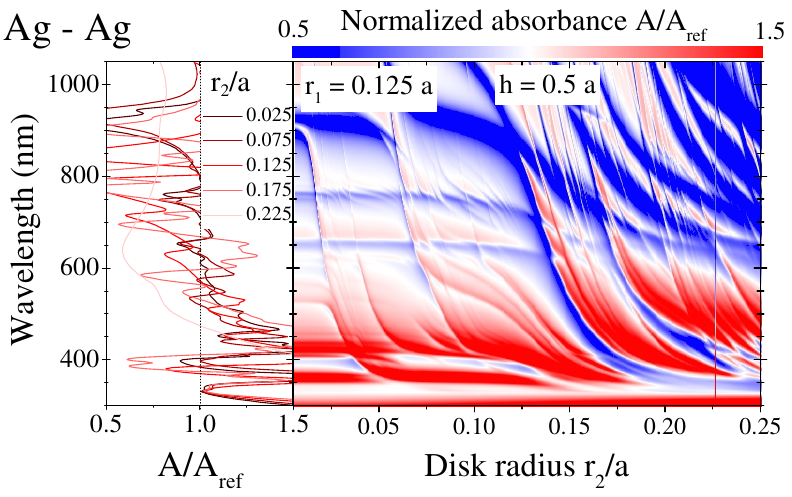}
\caption{Ag 2-type cells using different disk radii $r_2$ while keeping constant height $h=100$nm and $r_1=25$nm as in Fig.\ref{fig4}(b) for the overall device absorption giving the impression of a much stronger beneficial effect from this type of particle layer. See also table \ref{tab.AgSi2type}.}%
\label{fig45}
\end{figure}

Fig. \ref{fig4}(a) shows the normalized absorbance for an all Si 2-type layer where two diagonal disks have a fixed radius of 25nm (0.125 a) and height of $h=0.5a=100$nm. The size of the other two disks is changed to produce the contour plot which in comparison to the corresponding 1-type Si result from Fig. \ref{fig4}(a) readily shows an enhancement of the absorption in the active region over a broad spectral range. However, for smaller sizes the absorption in the blue range is further suppressed through absorption in the disks which yields a smaller, but still positive, enhancement of the short circuit current and the average absorption, resp. Fig. \ref{fig4}(c) shows the dependence for different disk heights on fixing both radii at $r_1=25$nm and $r_2=45$nm. The normalized absorbance is close to 1 for small disk heights, but increases significantly for taller disks. Furthermore, a decrease in performance can be seen for too high disks. This is quantified for several specific combinations in table \ref{tab.Si2type}. The corresponding spectra are plotted on the left hand side of the contour plots in the respective figures.

\begin{figure}%
\centering
\includegraphics[width=\textwidth]{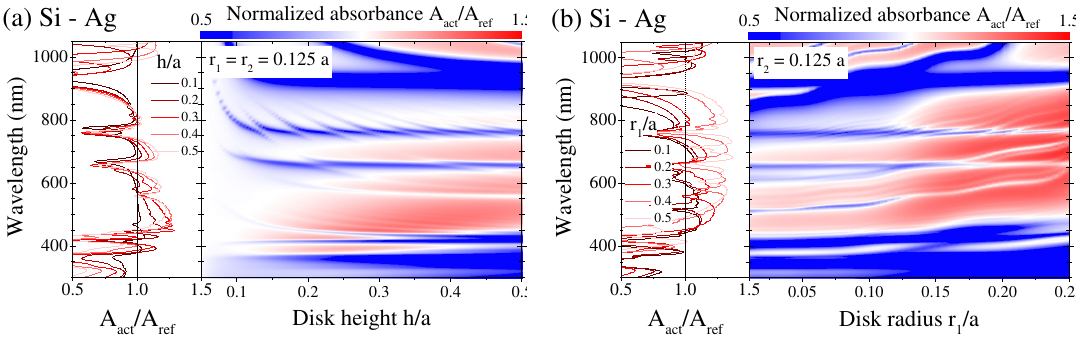}
\caption{2-type cell using Si and Ag disks (a) of equal size as a function of the disk height and (b) as a function of the radius of the Si disks $r_1$ for fixed radius $r_2=25$nm of the Ag disks.}%
\label{fig5}
\end{figure}

\begin{table}%
\centering
\begin{tabular}{c|c|c|c|c}
	Si $r_2$ & Ag $r_1$ & $h$ & $\eta_{\rm{Isc}}$ & $\eta_{\rm{Aav}}$ \\\hline
	 5 nm & 25 nm & 100nm &  -12\% & -15\% \\
	15 nm & 25 nm & 100nm &  -12\% & -15\% \\
	25 nm & 25 nm &  20nm &   -9\% & -11\% \\
	25 nm & 25 nm &  40nm &   -9\% & -11\% \\
	25 nm & 25 nm &  60nm &   -8\% & -12\% \\
	25 nm & 25 nm &  80nm &   -6\% & -12\% \\
	25 nm & 25 nm & 100nm &   -6\% & -16\% \\
	35 nm & 25 nm & 100nm &   +1\% &  -8\% \\
	45 nm & 25 nm & 100nm &  +11\% &  +1\% \\\hline\hline
	Ag $r_2$ & Ag $r_1$ & $h$ & $\eta_{\rm{Isc}}$ & $\eta_{\rm{Aav}}$ \\\hline
	25 nm & 25 nm & 100nm & -29\% & -29\% \\\hline\hline
	Ag $r_2$ & Ag $r_1$ & $h$ & $\eta_{\rm{Isc}}(A)$ & $\eta_{\rm{Aav}}(A)$ \\\hline
	25 nm & 25 nm & 100nm & +3\% & +11\%
\end{tabular}
\caption{Efficiency enhancement calculated from the short circuit current and average absorption relative to the reference cell for Si-Ag 2-type cells from Fig. \ref{fig5} for different particle sizes and heights. We include here the results from 1-type cells for both overall device absorption (Fig. \ref{fig45}) and absorption occurring exclusively inside the photo-active region (Fig. \ref{fig3}(b)) for comparison.} 
\label{tab.AgSi2type}
\end{table}

Similarly Fig. \ref{fig4}(b) and (d) show these results using Ag as the particle material. Here, too, spectral regions with an increase in the absorbance $A_{\rm act}$ can be seen. However, it is not strong enough to yield positive efficiency boosts (integrated over the whole spectrum) in the photo-active region. On the other hand, the overall device absorption shown in Fig. \ref{fig45} (comparing with Fig. \ref{fig4}(b)) yields an increase in absorbance and short circuit current, which underlines the importance to separate between the benefits of the particle layer and parasitic absorption effects arising from its employment. The results calculated from the overall absorbance $A$ lead to believe that this particle array has a positive effect on the device performance with an increase in both short circuit current $I_{\rm{sc}}$ and average absorption $A^{\rm{av}}_{\rm{act}}$. The opposite is found experimentally showing a considerable reduction in the internal quantum efficiency at lower wavelengths.\cite{Cortes-Juan2013} This is summarized in terms of the relative efficiency enhancement in table \ref{tab.AgSi2type}.

Fig. \ref{fig5}(a) shows 2-type particle arrays combining the two different materials Si and Ag as a function of disk height. These mixed unit cells demonstrate the spectral overlap of characteristic features from both 1-type cells (Fig. \ref{fig3}) or pure material 2-type cells (Fig. \ref{fig4}): the sharp resonance features (resulting in increased reflection or parasitic absorption) of the Ag particles and the significant, almost size-independent enhancement due to the Si particles. The light management improves in comparison to the pure Ag particle layer, but shows no further improvement with respect to the pure Si layer. In Fig. \ref{fig5}(b) the Ag disk radius is $r_2=25$nm and the Si disk radius is changed. An increase of the efficiency in the mixed 2-type unit cells can be seen only for Ag disks of smaller size than the Si disks, cp. table \ref{tab.AgSi2type}, and we conclude that employing Ag is detrimental to the absorbance in the active cell region for all configurations considered here.
While the effect of combining different materials is easily demonstrated with this theoretical method, it remains difficult to achieve in experiments.

\begin{figure}%
\centering
\includegraphics[width=0.45\textwidth]{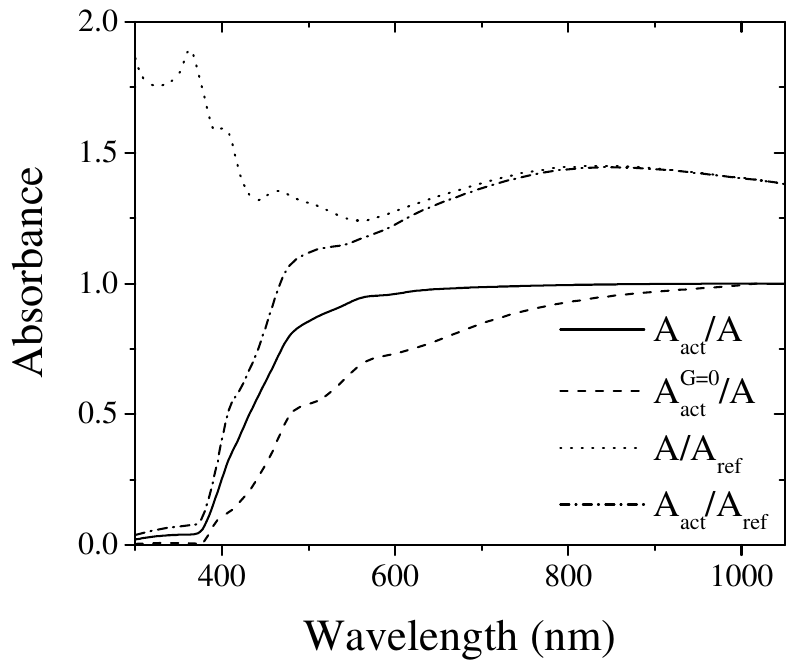}
\caption{Optical response of a 4-type cell made of  Si disks using four different disk radii, namely $r_1=15\textrm{nm },r_2=25\textrm{nm },r_3=35\textrm{nm },r_4=45$nm.}%
\label{fig7}%
\end{figure}

We furthermore consider 4-type particle layers, where we combine Si disk particles of four different disk radii in Fig. \ref{fig7}, namely $r_1=15\textrm{nm },r_2=25\textrm{nm },r_3=35\textrm{nm },r_4=45$nm and a periodicity of $a=400$nm. The efficiency enhancement in this example is $\eta_{\rm{Isc}}=40\%, \eta_{\rm{Aav}}=30\%$ within the active region. Despite the use of several smaller disks, this constitutes an increase in the efficiency compared to the 2-type cells which underlines the effect of boosting the absorption on providing a multitude of optical modes assisted by the particle layer. The increased difference between lowest order diffraction response and full calculation show the importance of diffracted light in the increase of absorbance. The increase in computational time for this example is moderate.

All the simulation results obtained here are without standard ARC coating from which an additional increase in the observed absorbance, resulting in a higher short circuit current, can be expected as it counterbalances reflection losses from back-scattering of the particle array. However, it cannot counterbalance losses due to increased absorption within the particle array.

\section{Conclusion}
In summary, we presented the absorbance within the photo-active region of a Si-wafer with a regular array of nanoparticles on top, using a combination of particles of both different size and materials in order to further enhance the short circuit current (up to $33\%$) and average absorption (up to $24\%$) relative to the bare reference cell. We assumed the absorption within the additional front layer of disk-shaped particles as parasitic for the enhancement of the photovoltaic effect and only considered the absorbance within the homogeneous photo-active region. Broadband efficiency enhancement in this active region is observed for pure dielectric (here Si) particle layers. Metal (here Ag) particle layers yield a much higher overall absorbance, but parasitic absorption losses - mostly due to ohmic heating - lead to a decrease in the internal quantum efficiency which can could be readily demonstrated in the presented theoretical calculations. Thus, only a partial improvement within the spectral range considered could be shown, but no broadband efficiency enhancement of the wavelength-independent short circuit current or average absorption was found.

Combining particles of different radii shows an increase in the absorbance compared to single type unit cells for a range of particle size combinations. Adjusting the height of the particle layer can further improve the benefit observed. A multi-type particle layer made of a single material of varying sizes can be fabricated with e.g. EBL techniques.
The creation of a spectral overlap of characteristic features in the optical properties of particle arrays was demonstrated by combining the different materials Si and Ag. Only a moderate increase in the computational effort results from the employment of multi-type unit cells and it remains a fast and reliable method.
Combining a larger number of differently sized particles improves the absorbance further due to a higher density of optical modes provided by the nanostructured front
surface layer.

\section{Acknowledgment}
The author thanks N. A. Mortensen and J. P. Connolly for stimulating discussions and the Deutsche Forschungsgemeinschaft (German Research Foundation) for financial support through a DFG research fellowship. The author would furthermore like to acknowledge the contribution of the European COST Action MP1406 MultiscaleSolar.
\end{document}